\documentclass[prb,twocolumn,aps,preprintnumbers,amsmath,amssymb]{revtex4}
\usepackage{graphicx}

\begin{document}
\preprint{}
\draft

\title{The entanglement beam splitter: a quantum-dot spin in a double-sided optical microcavity}

\author{C.Y.~Hu$^{1}$}\email{chengyong.hu@bristol.ac.uk}
\author{W.J.~Munro$^{2,3}$}
\author{J.L.~O'Brien$^1$}
\author{J.G.~Rarity$^1$}
\affiliation{$^1$Department of Electrical and Electronic Engineering, University of Bristol, University Walk, Bristol BS8 1TR, United Kingdom}
\affiliation{$^2$Hewlett-Packard Laboratories, Filton Road, Stoke Gifford, Bristol BS34 8QZ, United Kingdom}
\affiliation{$^3$National Institute of Informatics, 2-1-2 Hitotsubashi, Chiyoda-ku, Tokyo 101-8430, Japan}

\begin{abstract}

We propose an entanglement beam splitter (EBS) using a quantum-dot spin in a double-sided optical microcavity.
In contrast to the conventional optical  beam splitter, the EBS can directly split a photon-spin product state
into two constituent entangled states via transmission and reflection with high fidelity and high efficiency
(up to 100 percent). This device is based on giant optical circular birefringence induced by a single
spin as a result of cavity quantum electrodynamics and the spin selection rule of trion transition (Pauli blocking).
The EBS is robust and it is immune to the fine structure splitting in a realistic quantum dot.
This quantum device can be used for deterministically creating  photon-spin, photon-photon and spin-spin entanglement as well as
a single-shot quantum non-demolition measurement of a single spin. Therefore, the EBS  can find wide applications in
quantum information science and technology.

\end{abstract}

\date{\today}


\maketitle

\section{Introduction}

The optical beam splitter is a fundamental device in optics \cite{loudon03} and is widely used in quantum information science and technology,
such as quantum communications \cite{gisin02}, linear-optics quantum computation \cite{klm01, obrien07} (based on non-classical interference
of photons \cite{hom87}), and even quantum metrology \cite{giovannetti04}. The conventional beam splitter is simply a cube with two triangular
prisms glued together, or a plate of glass with a thin coating. The photon reflected or transmitted by such a beam splitter leaves no imprint
in the splitter itself, and the conventional beam splitter can not deterministically generate photonic entanglement \cite{fattal04}.
In this paper, we propose a beam splitter consisting of a singly charged quantum dot (QD) strongly coupled to a double-sided microcavity.
Such a structure can act as an entanglement beam splitter(EBS) which directly splits an initial product state of photon and spin into
two entangled states via transmission and reflection in a deterministic way.

In this paper, we introduce the transmission  and reflection operators for such an EBS as
\begin{equation}
\begin{split}
&\hat{t}=|R\rangle\langle R|\otimes
|\uparrow \rangle\langle \uparrow|+|L\rangle\langle L|\otimes |\downarrow \rangle\langle \downarrow| \\
&\hat{r}=|R\rangle\langle R|\otimes |\downarrow \rangle\langle \downarrow|
+|L\rangle\langle L|\otimes |\uparrow \rangle\langle \uparrow|,
\end{split} \label{ebs}
\end{equation}
where $|R\rangle$, $|L\rangle$ are right-circular and left-circular polarization states of photon and $|\uparrow \rangle$, $|\downarrow \rangle$
are the electron spin-up and spin-down states. We note that the transmission or reflection operator for our splitter is the same as the entanglement
filter recently demonstrated  by Okamoto et al  \cite{okamoto09}. This entanglement filter is constructed from partially polarizing beam splitters and can be used for the creation and purification of photon-photon entanglement probabilistically (with $1/16$ success probability)\cite{hofmann02}. However our EBS is in principle 100\% efficient in converting the input photon-spin product state $(|R\rangle + |L\rangle)(|\uparrow \rangle + |\downarrow \rangle)/2$ into the two constituent entangled states:
$(|R\rangle |\uparrow \rangle+|L\rangle|\downarrow \rangle)\sqrt{2}$ in the transmission port
and $(|R\rangle|\downarrow \rangle+|L\rangle|\uparrow \rangle)/\sqrt{2}$ in the reflection port.
Besides the photon-spin entanglement, we show that the EBS can easily be configured to transfer the quantum state between a photon and a spin by measuring
out the photon (spin) information in a suitable basis. Similarly by sending two photons in a product state into the EBS,
all combinations of photon outputs can be shown to be deterministically entangled once the spin is measured out. As a result,
the EBS can be widely applied in quantum information science and technology.

In our previous work, we introduced another photon-spin entangling gate \cite{hu08}. The current EBS and  the previous gate
are both deterministic for entanglement generation and are immune to the fine structure splitting in realistic semiconductor QDs.
The two quantum  devices are both based on the giant circular birefringence induced by a single spin strongly coupled
to an optical microcavity. The giant circular birefringence for such spin-cavity systems can manifest itself as the differences
in e.g., effective refractive index, phase, or reflection/transmission coefficients between the two circular polarizations.
The previous gate works in the reflection geometry and is constructed from a single-sided cavity with one mirror partially reflective and another mirror $100\%$ reflective. Such a spin-cavity system shows large phase difference between the uncoupled (cold) cavity and the coupled (hot) cavity, which induces the giant Faraday rotation and is exploited to build the gate. This gate is fragile because it demands that the reflectance for the uncoupled and the coupled cavity should be balanced to get high  fidelity.  However, the EBS works in reflection/transmission geometry, and is constructed from a double-sided cavity with both mirrors partially reflective.  Such a spin-cavity shows
large reflectance and transmittance difference between the uncoupled and the coupled cavity, which is exploited to build the EBS.
The larger the spin-cavity coupling strength, the higher the EBS fidelity. Therefore, the EBS is robust
and flexible compared to the previous gate.

The paper is organized as follows: In Sec. II, the photon-spin EBS is introduced, and we discuss the
efficiency and the fidelity to generate the photon-spin entanglement. We also discuss how the photon-spin entanglement can be
used for a single-shot quantum non-demolition measurement of a single spin. In Sec. III we show the EBS can be used to implement
a photon-spin quantum interface. In Sec. IV, we discuss another application of the EBS in entangling independent photons, and
show how remote spin entanglement can be also generated by the EBS. Finally, we present our conclusions and the outlook.

\section{Photon-spin entanglement beam splitter}

We consider a singly charged QD (e.g., self-assembled In(Ga)As QD,
GaAs interfacial QD, or semiconductor nanocrystal) placed at the antinode of an double-sided optical microcavity. For example, Fig. 1(a) shows a
micropillar microcavity where two GaAs/Al(Ga)As distributed Bragg reflectors (DBR) and transverse index guiding provide three-dimensional
confinement of light. In contrast to our previous work \cite{hu08}, the two DBRs are made partially reflective (double-sided), symmetric, and low loss
in order to achieve high on-resonance transmission. The cross section of the micropillar is made circular to make the cavity mode
degenerate for circularly polarized light.

\begin{figure}[ht]
\centering
\includegraphics* [bb= 104 202 500 625, clip, width=5cm,height=5cm]{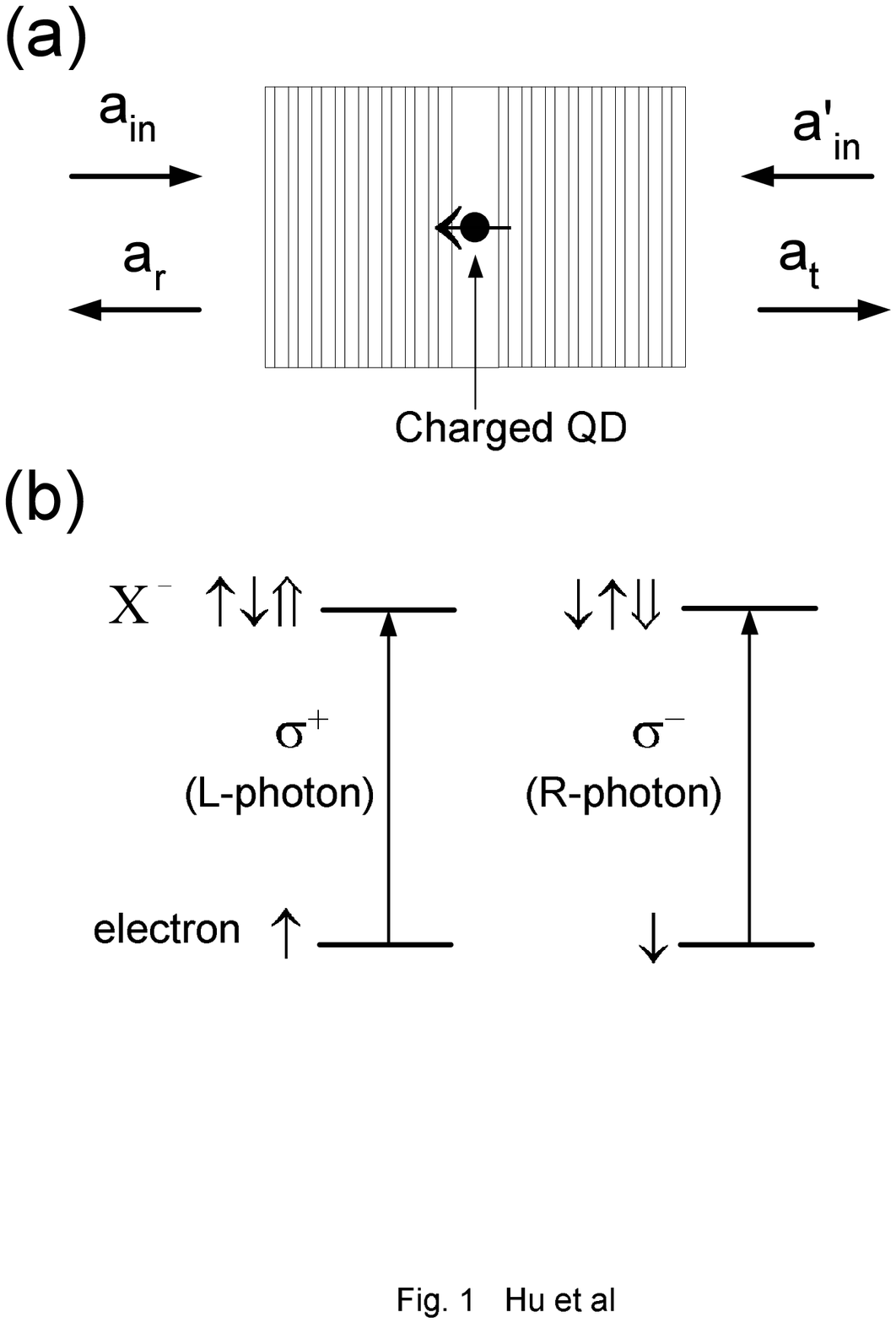}
\caption{(a) A charged QD inside a micropillar microcavity with circular cross section.
(b) Energy levels and spin selection rules for the optical transitions of $X^-$ (see text).} \label{fig1}
\end{figure}

Singly charged QDs show the optical resonance of trion $X^-$ (also called negatively-charged exciton)
which consists of two electrons bound to one hole \cite{warburton97}.
Due to the Pauli exclusion principle, $X^-$ shows spin-dependent
optical transitions [see Fig. 1(b)]\cite{hu98}: the left circularly polarized photon (indicated by  L-photon) only couples the electron in the
 spin state $|\uparrow\rangle$ to $X^-$ in the spin state $|\uparrow\downarrow\Uparrow\rangle$ with the two electron spins antiparallel; the right
circularly polarized photon (indicated by  R-photon) only couples the electron in the spin state $|\downarrow\rangle$ to $X^-$ in the
spin state $|\downarrow\uparrow\Downarrow\rangle$. Here $|\uparrow\rangle$  and $|\downarrow\rangle$ represent electron spin states $|\pm
\frac{1}{2}\rangle$, $|\Uparrow\rangle$ and $|\Downarrow\rangle$ represent heavy-hole spin states $|\pm\frac{3}{2}\rangle$.
This spin selection rule for $X^-$ is also called Pauli blocking \cite{warburton97, calarco03}, and
its imperfection due to the heavy-light hole mixing will be discussed later. For self-assembled QDs, the in-plane anisotropy
lifts the degeneracy of the bright neutral exciton, which is called fine structure splitting, via the anisotropic electron-hole exchange interactions, but induces no splitting for the charged exciton $X^-$ due to the quenched exchange interaction \cite{bayer02}. Therefore, the spin levels of $X^-$ are degenerate as shown in Fig. 1(b). However, the in-plane anisotropy could modify the hole states and thus affect the spin selection rule.
The light-hole sub-band and the split-off sub-band are energetically far apart from the heavy-hole sub-band and can be neglected.
The spin is quantized along the normal direction of the cavity, i.e., the propagation direction of the input (or output)
light.

For GaAs-based or InAs-based charged QDs, recent experiments have shown long electron spin coherence time ($T^e_2\sim\mu$s)
after suppressing the nuclear spin fluctuations \cite{petta05}, and long electron spin relaxation time ($T^e_1\sim$ms) \cite{kroutvar04}
due to the suppressed electron-phonon  and spin-orbit interactions in QDs.  These results indicate that QD-confined spin is one of the promising solid-state quantum-bit (qubit) systems.

The reflection and transmission coefficients of this $X^-$-cavity system can be investigated by solving the Heisenberg equations of
motion for the cavity field operator $\hat{a}$ and $X^-$ dipole operator $\sigma_-$, and  the input-output relations \cite{walls94}:
\begin{equation}
\begin{cases}
& \frac{d\hat{a}}{dt}=-\left[i(\omega_c-\omega)+\kappa+\frac{\kappa_s}{2}\right]\hat{a}-\text{g}\sigma_- \\
& ~~~~~~ -\sqrt{\kappa}\hat{a}_{in}-\sqrt{\kappa}\hat{a}^{\prime}_{in} +\hat{H}\\
& \frac{d\sigma_-}{dt}=-\left[i(\omega_{X^-}-\omega)+\frac{\gamma}{2}\right]\sigma_--\text{g}\sigma_z\hat{a}+\hat{G} \\
& \hat{a}_{r}=\hat{a}_{in}+\sqrt{\kappa}\hat{a} \\
& \hat{a}_{t}=\hat{a}^{\prime}_{in}+\sqrt{\kappa}\hat{a} \\
\end{cases}
\label{eq1}
\end{equation}
where $\omega$, $\omega_c$, and $\omega_{X^-}$ are the frequencies of the input photon, cavity mode, and $X^-$ transition, respectively.
g is the $X^-$-cavity coupling strength and $\gamma/2$ is the $X^-$ dipole decay rate. $\kappa$, $\kappa_s/2$ are  the cavity
field decay rate into the input/output modes, and the  leaky modes, respectively. The  background absorption
can also be included in $\kappa_s/2$. $\hat{H}$, $\hat{G}$ are the noise operators related to reservoirs.
$\hat{a}_{in}$, $\hat{a}^{\prime}_{in}$ and $\hat{a}_{r}$, $\hat{a}_{t}$ are the input and output field operators.

In the approximation of weak excitation where the charged QD is
predominantly in the ground state, we take $\langle \sigma_z\rangle \approx -1$. The reflection and transmission
coefficients in the steady state can be obtained
\begin{equation}
\begin{split}
& r(\omega)=1+t(\omega) \\
& t(\omega)=\frac{-\kappa[i(\omega_{X^-}-\omega)+\frac{\gamma}{2}]}{[i(\omega_{X^-}-\omega)+
\frac{\gamma}{2}][i(\omega_c-\omega)+\kappa+\frac{\kappa_s}{2}]+\text{g}^2}.
\end{split}
\label{eq2}
\end{equation}
The weak excitation approximation demands that the intracavity photon number should be less than
the critical photon number $n_0=\gamma^2/2g^2$ which measures the number of photons
in the cavity required to saturate the QD response \cite{kimble94}.
This condition is satisfied if the input beam is a train of single photons with the photon rate
less than $n_0/\tau$, where $\tau$ is the cavity lifetime. This situation
is what we consider in this work \cite{exp1}. Moreover, for the high fidelity operation of the EBS,
the spectral width of single photons should be much smaller than the linewidth of the cavity mode, i.e.,
the coherence length of single photons should be much longer than the cavity lifetime. This means
the photon is never completely localized in the cavity, therefore, the charged QD almost remains in the ground
state all the time.  Although single photons as the input are discussed in this work and
our previous work \cite{hu08} (as single photons are robust), we have to emphasize here
that all discussions can be extended to a weak laser as the input as long as the weak
excitation condition is satisfied. This can be done by replacing the single photon state
with the coherent state. Details shall be discussed elsewhere \cite{hu09}.

Instead of the dispersive interaction \cite{raimond01}, we consider here the resonant interaction
with $\omega_c=\omega_{X^-}=\omega_0$. By taking $\text{g}=0$,
we get the reflection and transmission coefficients $r_0(\omega)$, $t_0(\omega)$
for a uncoupled cavity (or cold cavity) where the QD does not couple to the cavity \cite{exp3}:
\begin{equation}
\begin{split}
& r_0(\omega)=\frac{i(\omega_0-\omega)+\frac{\kappa_s}{2}}{i(\omega_0-\omega)+\kappa+\frac{\kappa_s}{2}} \\
& t_0(\omega)=\frac{-\kappa}{i(\omega_0-\omega)+\kappa+\frac{\kappa_s}{2}}.
\end{split}
\label{eq3}
\end{equation}

The reflection and transmission spectra versus the detuning $\omega-\omega_0$ are presented in Fig. 2(a) for different
coupling strength $\text{g}$. With increasing $\text{g}$ (e.g. by reducing the modal volume or increasing the $X^-$ oscillator strength),
the cavity mode splits into two peaks due to the quantum interference  in the \textquotedblleft one-dimensional
atom\textquotedblright regime \cite{waks06} with $\kappa < 4\text{g}^2/\kappa < \gamma$  (which has
been demonstrated recently \cite{englund07}), and the vacuum Rabi splitting in the strong coupling regime \cite{reithmaier04}
with $\text{g} > (\kappa, \gamma)$.
We can see  that the transmittance or reflectance are different between the uncoupled cavity with $\text{g}=0$ and the coupled
cavity (or hot cavity) with $\text{g}\neq 0$.

\begin{figure}[ht]
\centering
\includegraphics* [bb= 91 459 496 751, clip, width=8.5cm, height=8cm]{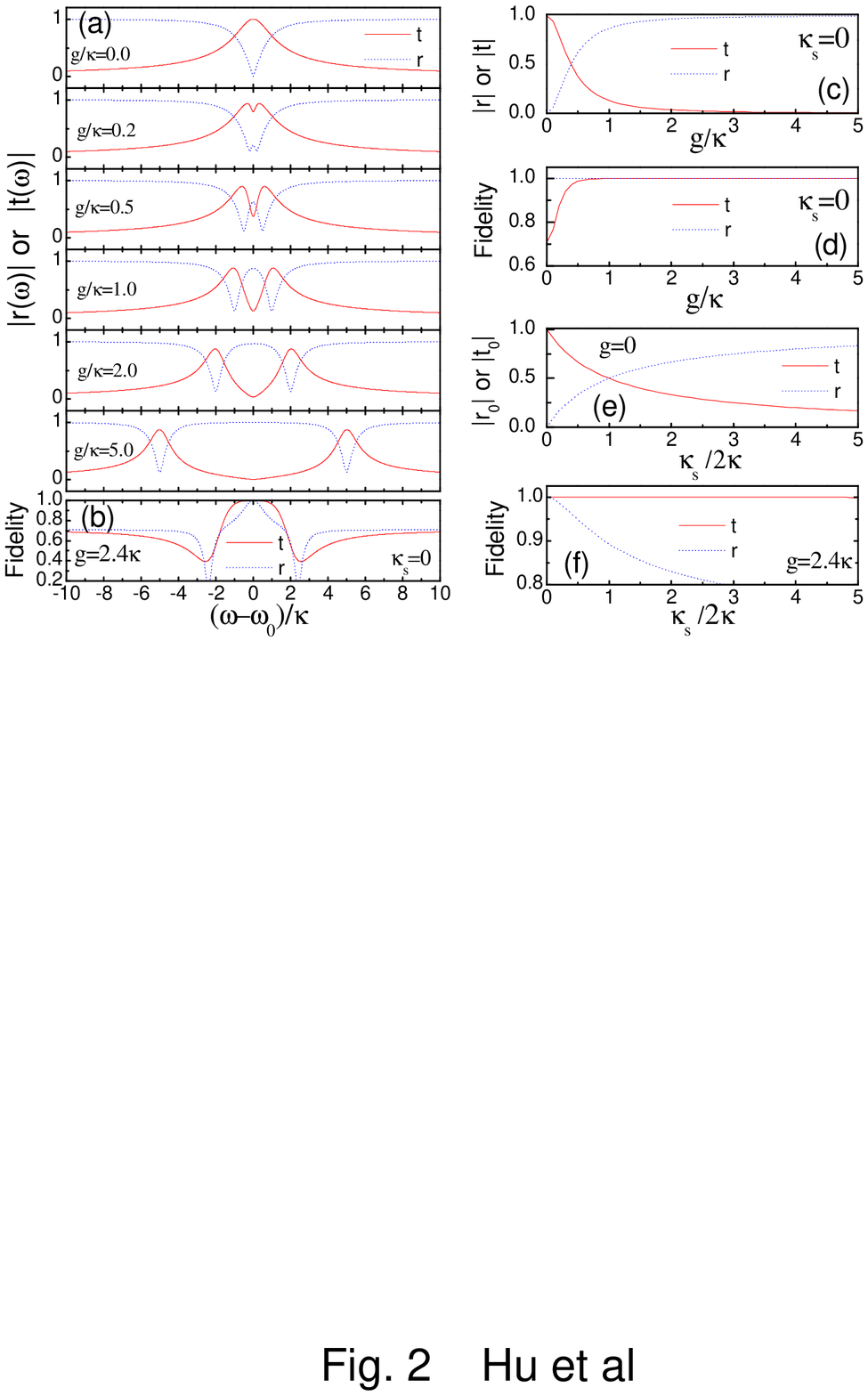}
\caption{(Color online) (a) Calculated transmission (solid curves)  and reflection (dotted curves) spectra vs the normalized frequency detuning
for different coupling strength $\text{g}$. (b) The entanglement fidelity vs the frequency detuning in the strong coupling regime
($\text{g}=2.4\kappa$ is taken). Solid curve for $F^t(\omega)$ and dotted curve for $F^r(\omega)$.
(c) Transmittance $|t(\omega_0)|$ (solid curve)  and  reflectance $|r(\omega_0)|$ (dotted curve) vs the normalized
coupling strength. (d) The  entanglement fidelity  vs the normalized coupling strength.
(e) Transmittance $|t_0(\omega_0)|$ (solid curve)  and  reflectance $|r_0(\omega_0)|$ (dotted curve) vs the normalized side leakage rate.
(f) The  entanglement fidelity  vs the normalized side leakage rate. Solid curve for $F^t(\omega_0)$ and dotted curve for $F^r(\omega_0)$.
$\omega_c=\omega_{X^-}=\omega_0$ and $\gamma=0.1\kappa$ are taken for (a)-(f), and $\kappa_s=0$ for (a)-(d). } \label{fig2}
\end{figure}

If the spin lies in the state $|\uparrow\rangle$, the L-photon feels a coupled cavity with  reflectance $|r(\omega)|$
and the transmittance $|t(\omega)|$, whereas the R-photon  feels the uncoupled cavity with the reflectance $|r_0(\omega)|$
and transmittance $|t_0(\omega)|$; Conversely, if the spin lies in the  state $|\downarrow\rangle$, the R-photon feels a
coupled cavity, whereas the L-photon feels the uncoupled cavity. The difference in transmission and reflection, or even in
the phase \cite{turchette95, hu08}, between the right and
left circular polarizations are all called the giant circular birefringence induced by a single spin. This enables us to make a
quantum device - an entanglement beam splitter. For any quantum input we can define the transmission operator as
\begin{equation}
\begin{split}
\hat{t}(\omega)=& t_0(\omega)(|R\rangle\langle R|\otimes |\uparrow \rangle\langle \uparrow|+|L\rangle\langle L|\otimes
|\downarrow \rangle\langle \downarrow|)\\
& +t(\omega)(|R\rangle\langle R|\otimes |\downarrow \rangle\langle \downarrow|+|L\rangle\langle L|\otimes |\uparrow \rangle\langle \uparrow|),
\end{split}
\label{eq4}
\end{equation}
and the reflection operator as
\begin{equation}
\begin{split}
\hat{r}(\omega)=& r_0(\omega)(|R\rangle\langle R|\otimes |\uparrow \rangle\langle \uparrow|+|L\rangle\langle L|\otimes
|\downarrow \rangle\langle \downarrow|)\\
& +r(\omega)(|R\rangle\langle R|\otimes |\downarrow \rangle\langle \downarrow|+|L\rangle\langle L|\otimes |\uparrow \rangle\langle \uparrow|),
\end{split}
\label{eq5}
\end{equation}
where $r_0(\omega)$, $t_0(\omega)$ and $r(\omega)$,  $t(\omega)$ are the reflection and transmission coefficients
of the uncoupled  and coupled cavity, respectively. Both operators include the contributions from the uncoupled
and coupled cavity.

In the strong coupling regime $\text{g} > (\kappa, \gamma)$ and in the central frequency regime $|\omega-\omega_0|< \kappa$,
we have $|t(\omega)|\rightarrow 0$ and $|t_0(\omega)|\neq 0$ [see Fig. 2(a)], thus the transmission operator
can be simplified as
\begin{equation}
\hat{t}(\omega)\simeq t_0(\omega)(|R\rangle\langle R|\otimes |\uparrow \rangle\langle \uparrow|+|L\rangle\langle L|\otimes
|\downarrow \rangle\langle \downarrow|),
\label{eq6}
\end{equation}
including the contribution from the uncoupled cavity only.
Under the same conditions, we get $|r(\omega)|\rightarrow 1$. If the side leakage is small,
i.e., $\kappa_s \ll \kappa$, we have $|r_0(\omega)| \rightarrow 0$ at $\omega \simeq \omega_0$ [see Eq.(3)].
So the reflection operator can be simplified as
\begin{equation}
\hat{r}(\omega)\simeq r(\omega)(|R\rangle\langle R|\otimes |\downarrow \rangle\langle \downarrow|+|L\rangle\langle L|\otimes
|\uparrow \rangle\langle \uparrow|),
\label{eq7}
\end{equation}
including the contribution from the coupled cavity only.

Thus we get the EBS transmission and reflection operators defined in Eq. (\ref{ebs}): $\hat{t}(\omega)\simeq t_0(\omega)\hat{t}$
and $\hat{r}(\omega)\simeq r(\omega)\hat{r}$ where the coefficients
determine the EBS efficiency.
As $|t_0(\omega)|\rightarrow 1$ and $|r(\omega)|\rightarrow 1$ when $\omega \simeq \omega_0$, the EBS can be made
$100\%$ efficient by optimizing the pillar cavity (e.g., suppress the side leakage and other loss \cite{reitzenstein07}).

\begin{figure}[ht]
\centering
\includegraphics* [bb= 134 550 488 661, clip, width=8cm,height=2.5cm]{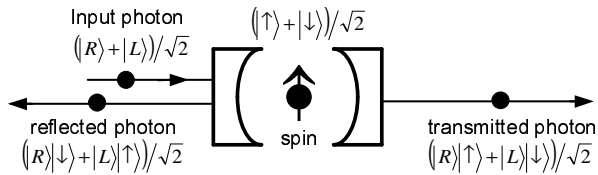}
\caption{The proposed photon-spin entanglement beam splitter. A photon-spin product state can be split into
two constituent photon-spin entangled states: one in the transmission port and another in the reflection port. Note that
the single photon is either transmitted or reflected, so this entanglement beam splitter is deterministic. } \label{fig3}
\end{figure}

The EBS can directly split a photon-spin product state into two constituent entangled states
via transmission and reflection (see Fig. 3). However, the entanglement fidelity is different between transmission and reflection.
We can define the amplitude entanglement fidelity for the transmission and reflection operators as
\begin{equation}
\begin{split}
&F^t(\omega)=|t_0(\omega)|/\sqrt{|t_0(\omega)|^2+|t(\omega)|^2} \\
&F^r(\omega)=|r(\omega)|/\sqrt{|r_0(\omega)|^2+|r(\omega)|^2}.
\end{split} \label{fid}
\end{equation}
As shown in Fig. 2(b) and Fig. 2(d), near-unity entanglement fidelity in transmission (solid curve) can be achieved
within a small frequency window $|\omega-\omega_0|<\kappa$ in the strong coupling regime $\text{g} > (\kappa, \gamma)$.
The strongly coupled QD-cavity has been demonstrated in various microcavities and nanocavities
recently  \cite{reithmaier04}, and $\text{g}/\kappa=2.4$ can be achieved for the In(Ga)As QD-cavity
system \cite{reithmaier04, reitzenstein07, exp2}. Experimentally $\gamma$ is about several $\mu$eV, so we take $\gamma=0.1\kappa$.
Our calculations are based on these experimental values. High fidelity in reflection occurs
in a much narrower bandwidth with unity achieved only at $\omega=\omega_0$ [see Fig. 2(b), dotted curve], and
$F^r(\omega_0)$ shows no dependence on $\text{g}$ [see Fig. 2(d), dotted curve] due to $|r_0(\omega_0)|=0$ when $\kappa_s\ll \kappa$,
suggesting we could make high-fidelity entanglement even in the weakly coupling regime.
However, a practical pillar microcavity has side leakage which reduces the entanglement fidelity [see Fig. 2(f)].  $F^t(\omega_0)$
shows a weak dependence on  the side leakage (solid curve), but $F^r(\omega_0)$ decreases rapidly (dotted curve)
because $|r_0(\omega_0)|\neq 0$ when $\kappa_s$ is significant.

The spin selection rule is not perfect for a realistic QD due to the heavy-light hole mixing. This can reduce the
entanglement fidelity by a few percent as the hole mixing in the valence band is in the order of a few percent \cite{bester03, calarco03}
[e.g., for self-assembled In(Ga)As QDs]. The hole mixing  could  be reduced by engineering the shape and size of QDs or choosing
different types of QDs.

The exciton dephasing can also reduce the entanglement fidelity by amount of $[1-\exp{(-\tau/T_2)}]$, where
$\tau$ is the cavity photon lifetime and $T_2$ is the exciton coherence time. Two kinds of dephasing
processes should be mentioned here: the optical dephasing and the spin dephasing of $X^-$. It is well-known that the optical coherence time
of excitons in self-assembled In(Ga)As QDs can be as long as several hundred picoseconds \cite{borri01}, which is ten times longer than the cavity photon lifetime (around tens of picoseconds in the strong coupling regime for cavity Q-factor of $10^4-10^5$). So the optical dephasing can only slightly reduce the entanglement fidelity by less than $10\%$. The spin dephasing
of the $X^-$ is mainly due to the hole-spin dephasing. In the absence of the contact hyperfine interaction that happens for holes \cite{heiss07},
the QD-hole spin is expected to have long coherence time, and $T^h_2>100$ ns has been reported recently \cite{brunner09}. This spin
coherence time is at least three orders of magnitude longer than the cavity photon lifetime, so the spin dephasing of the $X^-$ can be safely neglected
in our considerations.

The photon-spin entanglement enables us to make an ideal quantum measurement of the single spin
by measuring the helicity of the transmitted or reflected photon.
Although the resonant interaction is considered here, the weak excitation condition
implies the real excitation of $X^-$ transition is negligible. As a result, the spin disturbance by
the input photon is small. This optical spin-detection method, as well as another way based on the giant
Faraday rotation \cite{hu08},  is thus a single-shot quantum non-demolition measurement (QND) \cite{grangier98}.

The QD-spin eigenstate can be prepared, for example, by optical spin pumping and cooling \cite{atature06}.
The proposed single-shot QND measurement could also be used to prepare/cool the spin state via single-photon measurement \cite{liu05},
or via quantum Zeno effect \cite{misra77}.
The spin superposition state can be made from the eigenstates by performing single spin rotations using nanosecond ESR pulses
or picosecond optical pulses as reported recently\cite{berezovsky08}.

\begin{figure}[ht]
\centering
\includegraphics* [bb= 142 471 444 671, clip, width=6cm, height=4.5cm]{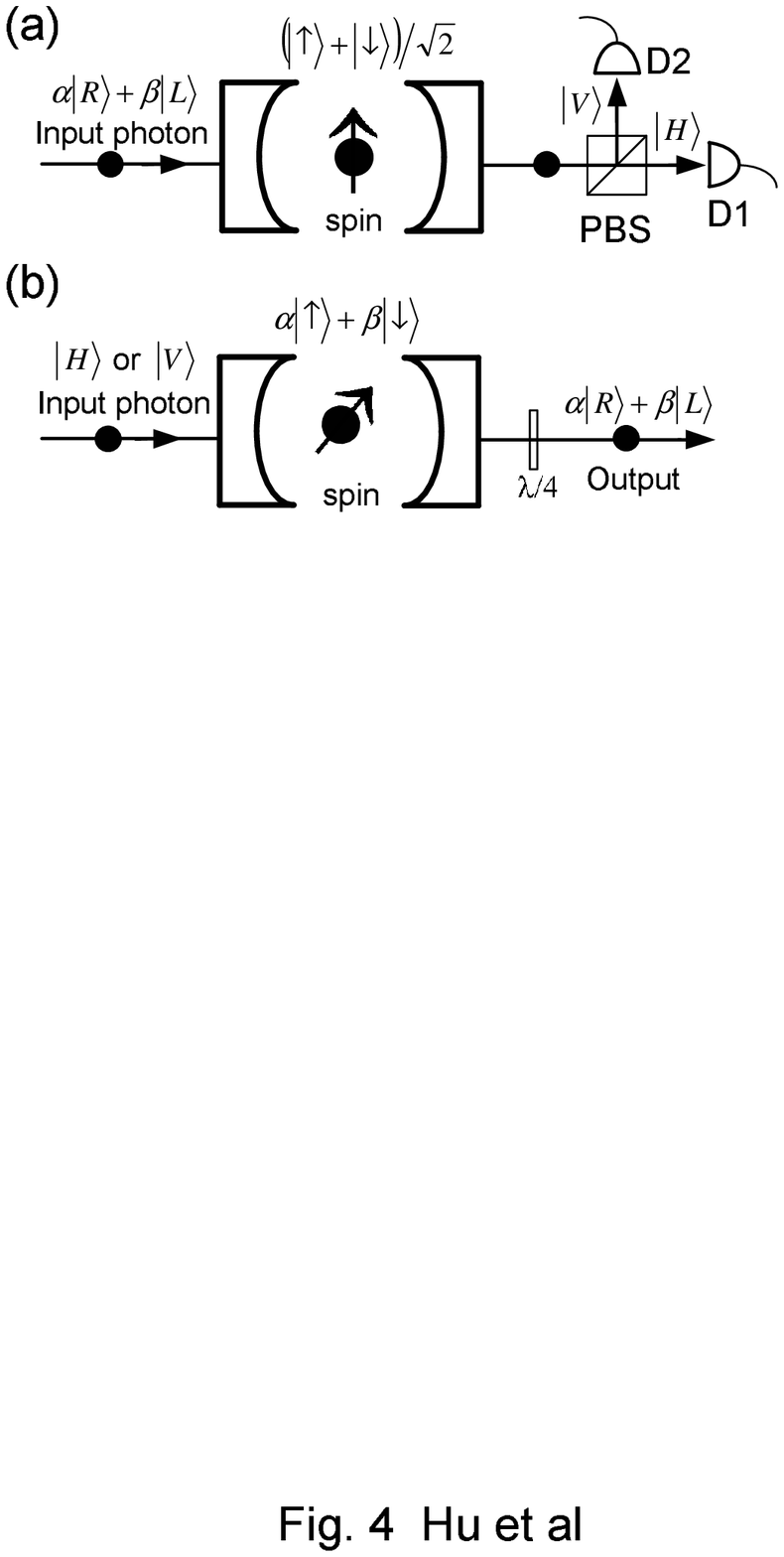}
\caption{Schematic diagram of a photon-spin interface. It also works in the  reflection geometry (not shown here).
(a) State transfer from a photon to a spin. (b) State transfer from a spin to a photon. PBS (polarizing beam splitter),
D1 and D2 (photon detectors), and $\lambda/4$ (quarter-wave plate).} \label{fig4}
\end{figure}

\section{Photon-spin quantum interface}

We could use the created photon-spin entanglement as a resource to teleport the state from a photon to a spin \cite{bennett93},
thus making a quantum interface \cite{yao05} which is a crucial component for quantum networks \cite{cirac97}.
However, we can do it directly using our EBS.
In Fig. 4(a), a photon in an arbitrary state $|\psi^{ph}\rangle=\alpha|R\rangle+\beta|L\rangle$ is sent to the cavity with
the spin prepared in the state $|\psi^{s}\rangle=(|\uparrow\rangle + |\downarrow\rangle)/\sqrt{2}$.
After transmission, the photon and the spin  become entangled, i.e,
\begin{equation}
|\psi^{ph}\rangle \otimes |\psi^{s}\rangle
\xrightarrow{\hat{t}(\omega)} \frac{t_0(\omega)}{\sqrt{2}} \left(\alpha |R\rangle |\uparrow\rangle+\beta |L\rangle|\downarrow\rangle\right).
\label{ps1}
\end{equation}
By applying a Hadamard gate on the photon state  using a polarizing beam splitter, we obtain a spin state
$|\Phi^s\rangle=\alpha|\uparrow\rangle\pm\beta|\downarrow\rangle$ on detecting a photon in the
$|H\rangle$ or $|V\rangle$ state. Therefore, the photon state is transferred to the spin.
The storage time is limited by the spin coherence time ($T^e_2 \sim \mu$s) \cite{petta05}.

In Fig. 4(b), a photon in the polarization state $|\psi^{ph}\rangle=(|R\rangle+|L\rangle)/\sqrt{2}$ is sent to the cavity with the
spin in an arbitrary state $|\psi^{s}\rangle=\alpha |\uparrow\rangle+\beta |\downarrow\rangle$.
After transmission, the state transforms to the same state as Eq. (\ref{ps1}).
After applying a Hadamard gate on the  spin (e.g., using a $\pi/2$ microwave or optical pulse \cite{berezovsky08}),
the spin is detected in the $|\uparrow\rangle$ and $|\downarrow\rangle$ basis by  the
QND measurement as discussed above. Depending on the measured spin states, the photon
is then projected in the state $|\Phi^{ph}\rangle=\alpha|R\rangle\pm\beta|L\rangle$, and the spin state is transferred
to the photon.

The state transfer can also be performed if the photon is reflected. Therefore, the total success probability is
$(|t_0(\omega)|^2+|r(\omega)|^2)/2$, which can reach $100\%$ if the cavity is optimized and  $\omega \simeq \omega_0$.
The state transfer fidelity is limited by the entanglement fidelity discussed in Sec. II.

\begin{figure}[ht]
\centering
\includegraphics* [bb= 76 346 537 671, clip, width=8cm, height=6cm]{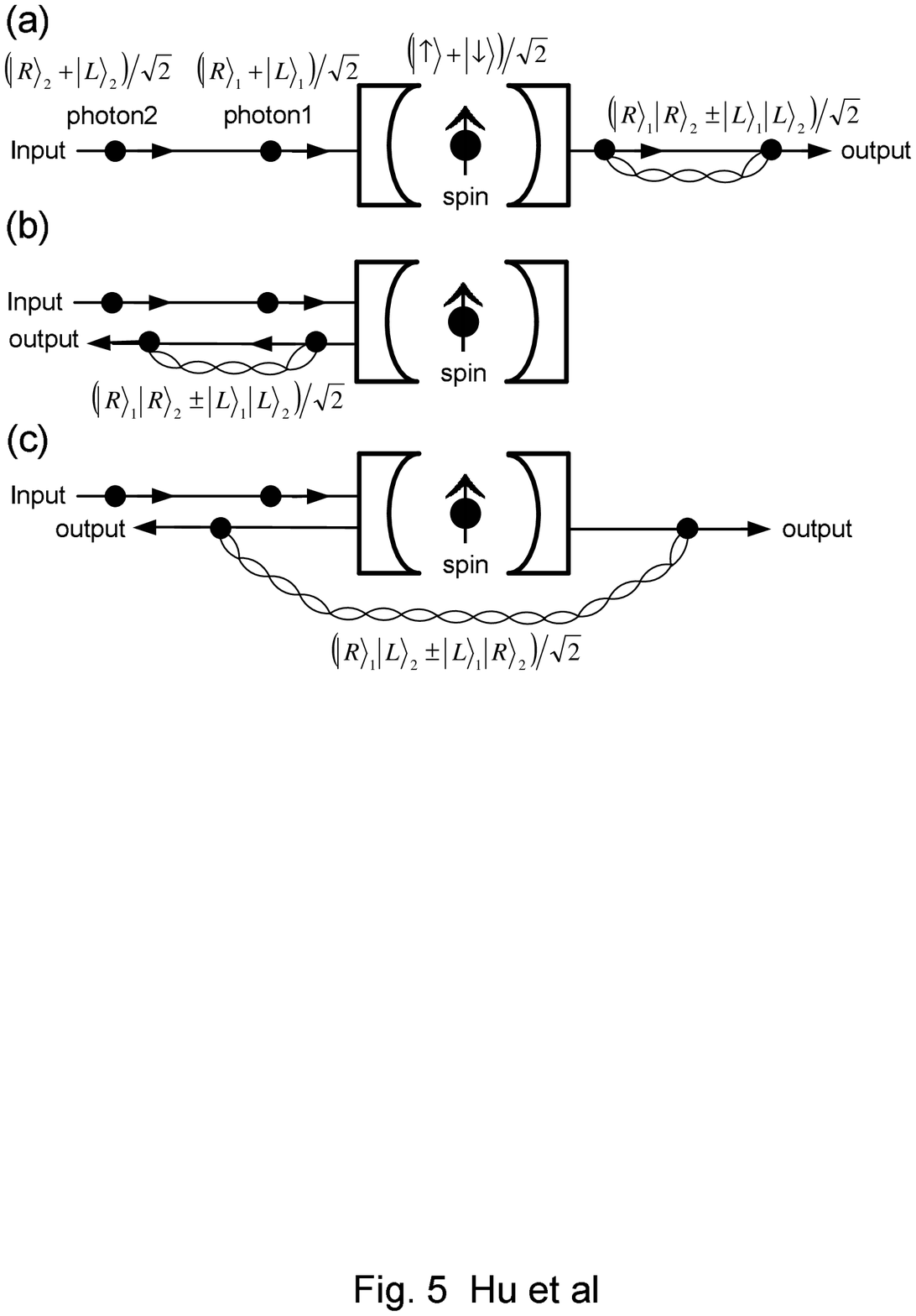}
\caption{An EBS-based scheme to entangle independent photons by measuring the spin states. The two photons can be (a) both transmitted, (b) both
reflected, or (c) one transmitted and another reflected. } \label{fig5}
\end{figure}

\section{Entangling independent photons}

Another application of the EBS is to create polarization entanglement between independent
photons (see Fig. 5). Photon 1 and photon 2 are both prepared in the superposition state
$|\psi^{ph}\rangle_{1,2}=(|R\rangle_{1,2}+|L\rangle_{1.2})/\sqrt{2}$ and sent to the cavity in sequence.
The spin is prepared in the state $|\psi^{s}\rangle=(|\uparrow\rangle+|\downarrow\rangle)/\sqrt{2}$.

If both photons are transmitted [see Fig. 5(a)] with a possibility of $25\%$, the state transformation is
\begin{equation}
\begin{split}
|\psi^{ph}\rangle_1 & \otimes |\psi^{ph}\rangle_2 \otimes |\psi^{s}\rangle \xrightarrow{\hat{t}(\omega)} \frac{t_0(\omega_1)t_0(\omega_2)}{2}\\
& \times \left(|R\rangle_1|R\rangle_2|\uparrow\rangle+|L\rangle_1|L\rangle_2|\downarrow\rangle \right)/\sqrt{2}.
\end{split}
\label{pe1}
\end{equation}

After applying a Hadamard gate on the electron spin \cite{berezovsky08},
the spin is detected in the $|\uparrow\rangle$ and $|\downarrow\rangle$ basis by the QND measurement using
another photon in H or V polarization. Depending on the measured spin states,
we get two Bell  states $\Phi_{12}^{\pm}=(|R\rangle_1|R\rangle_2 \pm |L\rangle_1|L\rangle_2)/\sqrt{2}$.

Similarly, if the two photons are both reflected [see Fig. 5(b)]with a possibility of $25\%$, the Bell states $\Phi_{12}^{\pm}$
can be obtained again. If one photon is transmitted and another reflected [see Fig. 5(c)] with a possibility of $50\%$, we get
another two Bell states
$\Psi_{12}^{\pm}=(|R\rangle_1|L\rangle_2 \pm |L\rangle_1|R\rangle_2)/\sqrt{2}$.
So the full set of Bell states can be created.
This EBS-based entanglement scheme is $100\%$ efficient
if the cavity is optimized and  $\omega_{1,2} \simeq \omega_0$ [see Fig.2(c) and (e)].
The time interval  between two photons should be much shorter than
the spin coherence time ($T^e_2 \sim \mu$s) to achieve high entanglement fidelity \cite{hu08}, but it should be
long enough to make the weak excitation approximation valid \cite{exp1}.
Note that this deterministic scheme for photon entanglement generation is immune to the fine structure
splitting in realistic semiconductor QDs \cite{bayer02} as discussed in Sec. II, in contrast to another scheme based on
the biexciton cascade emissions \cite{stevenson06} where the fine structure splitting should be suppressed to get high entanglement
fidelity (this is a hard task).

This scheme can be extended to deterministically generate multi-photon entanglement \cite{devitt07},
such as Greenberger-Horne-Zeilinger(GHZ) states \cite{greenberger90} or cluster states \cite{briegel01},
by repeating the above Bell-state creation procedure to
increase the size. An alternative way to create multi-qubit entanglement is entangling in different degrees of freedom
with no need to increase the number of photons. This type of entanglement is called hyper-entanglement\cite{kwiat97}.
If we take the spatial modes (different paths) in Fig. 5 into account, we actually create the two-photon three-qubit states entangled
in polarization and spatial modes.

Besides the scalable photon-spin and photon-photon entanglement, the EBS can be also used to deterministically create  remote entanglement
between spins in different cavities via a single photon \cite{hu08} which in fact acts as a quantum bus \cite{spiller06}.

\section{Conclusions and outlook}

We have proposed an entanglement beam splitter which can directly split a product state into two constituent entangled states with high
fidelity and $100\%$ efficiency using a QD spin strongly coupled to an optical microcavity. This device is robust and  is immune to
the fine structure splitting in a realistic QD. This entanglement beam splitter can be used to
deterministically create photon-spin, photon-photon, and spin-spin entanglement. This high-fidelity entanglement would find wide applications in quantum communications such as  quantum cryptography and quantum teleportation \cite{gisin02}. Moreover, this entanglement is essential to implement
a quantum bus\cite{spiller06}, quantum interface \cite{yao05}, quantum memories and  quantum repeaters \cite{briegel98}, all of which are critical building blocks for quantum networks \cite{cirac97}. The high-order multiparticle entanglement could be used for entanglement-enhanced quantum measurement \cite{giovannetti04}, or cluster-state based quantum computing \cite{raussendorf01, nielsen06}. The single-shot QND measurement
could be used to prepare or cool the single spin state.

The entanglement beam splitter can also work as an active device such as a polarization-controlled single photon source driven by the electron spin dynamics. This source benefits from cavity quantum electrodynamics, therefore it has high high quantum efficiency \cite{cui05} and time-bandwidth limited photon pulses \cite{santori02}. On the other hand, techniques for manipulating single photons are well developed, and significant progress on fast QD-spin cooling and manipulating has been made recently \cite{atature06, berezovsky08}. Together with this work and our
previous work \cite{hu08}, we believe that the QD spin-cavity systems are quite promising for solid-state quantum networks and scalable quantum computing.

It's worthy to point out here that the entanglement beam splitter can be also made from a QD-hole spin in a double-sided microcavity because the positively charged exciton is governed by the same spin selection rule as the negatively charged exciton \cite{heiss07, brunner09}.
Recent experiments have demonstrated that the QD-confined hole has long spin relaxation time ($T^h_1\sim 1$ ms) \cite{heiss07} and spin coherence time ($T^h_2>100$ ns)\cite{brunner09} due to the suppressed electron-phonon interaction and the lack of hole-nuclear hyperfine interaction. Moreover,
the cooling and fast coherent control of hole-spin states have been reported recently \cite{heiss07, ramsay08, brunner09}.
Therefore, this type of entanglement beam splitter based on the hole spin is also promising.

\section*{ACKNOWLEDGMENTS}

C.Y.H. thanks M.S. Kim, M. Atat\"{u}re, X.Q. Zhou, S. Bose, and S. Popescu for helpful discussions.
J.L.O'B. and J.G.R. each acknowledge support from the Royal Society.
This work is partly funded by EPSRC-GB IRC in Quantum Information Processing, QAP (Contract No. EU IST015848), and MEXT from Japan.


\begin{references}
\bibitem{exp1}By taking the parameters used in this paper, i.e, $g/\kappa=2.4$ and $\gamma/\kappa=0.1$,
 we get the critical photon number $n_0=8.7\times 10^{-4}$. If taking $\tau=100$ ps
for a microcavity with $Q=3\times10^4$ and $\lambda_c=1~\mu$m, the single photon rate should be less than $8.7$ MHz to make the weak excitation approximation valid. If the mode-mismatch is included, the single photon rate can be slightly higher.
\bibitem{exp3}Experimentally we can measure the side leakage rate $\kappa_s/\kappa$ by measuring the ratio $r_0(\omega_0)/t_0(\omega_0)$.
\bibitem{exp2}$\text{g}$ and $\kappa$ can be controlled independently as $\text{g}$ is determined by the
$X^-$ oscillator strength and the cavity modal volume, while $\kappa$ by the cavity quality factor only.
\end{references}
\end{document}